# To make a glass – avoid the crystal

Thomas Palberg[1], Eckhard Bartsch[2], Richard Beyer[1], Maximilian Hofmann[1], Nina Lorenz[1,3], Janina Marquis[1], Ran Niu[1,4] and Tsuneo Okubo[5]

Abstract

**Colloidal model systems allow for a flexible tuning of particle sizes, particle spacings and mutual interactions at constant temperature. Colloidal suspensions typically crystallize as soon as the interactions get sufficiently strong and long-ranged. Several strategies have been successfully applied to avoid crystallization and instead produce colloidal glasses. Most of these amorphous solids are formed at high particle concentrations. This paper shortly reviews experimental attempts to produce amorphous colloidal solids using strategies based on topological, thermodynamic and kinetic considerations. We complement this overview by introducing a (transient) amorphous solid forming in a thoroughly deionized aqueous suspension of highly charged spheres at low salt concentration and very low volume fractions.**

Key words: colloidal model systems, colloidal glasses, structural glasses, light scattering, crystal nucleation, polydispersity, eutectic phases, frustration, dynamic arrest,

✉    Thomas Palberg
     palberg@uni-mainz.de

[1] Institute of Physics, Johannes Gutenberg University, D-55099 Mainz, Germany

[2] Institute of Physical Chemistry, Albert-Ludwigs-University, D-79104 Freiburg, Germany

[3] present address: Sentronics Metrology, Mannheim, Germany

[4] on leave from: Changchun Institute of Applied Chemistry, Chinese Academy of Sciences, Changchun 130022, China

[5] Institute for Colloidal Organization, Hatoyama 3-1-112 Uji, Kyoto 611-0012, Japan

**Introduction**

Colloidal suspensions are well known experimental model systems to study the structural and dynamic properties of condensed matter [1, 2, 3, 4]. Typically, spherical particles of sub-micron size suspended in a liquid carrier medium are studied as a model for simple atomic solids and liquids [5]. Like their atomic counterparts, colloids show a first order phase transition from the fluid to the crystalline state, once their number density, $n$, or volume fraction, $\Phi = n(4/3)\pi a^3$, exceeds a critical value (where a denotes the particle radius). For systems with hard sphere (HS) interactions, this has been suggested from simulations by Hoover and Ree [6], and later confirmed experimentally by Pusey and van Megen [7]. Interestingly, the latter authors also observed an amorphous solid at large density, termed colloidal glass [8]. This study initiated a vibrant interest in colloids as model atoms and led to a large number of fascinating experiments on colloidal solids [9, 10, 11, 12, 13, 14, 15, 16, 17, 18, 19, 20, 21, 22, 23, 24, 25]. The freezing transition of HS is entropy driven and hence, it depends only on the volume fraction of particles. In charged sphere suspensions, phase transitions depend on the strength and range of the screened Coulomb repulsion [26, 27]. With a few interesting exceptions [28, 29, 30], melting, crystallization and vitrification are typically not controlled by temperature. They occur isothermally with the suspending medium acting as an effective heat bath [31]. Particle motion is diffusive with relevant time scales condensed to the conveniently accessible range of microseconds to hours or sometimes days [32]. Most samples are transparent with particle sizes and particle distances in the range of the wavelength of visible light. This allows for using microscopy and light scattering to investigate colloidal crystals and glasses [1, 2, 33, 34, 35]. Due to the small number density colloidal solids show a very low shear rigidity with moduli on the order of $G \approx 1Pa$ [36, 37]. Therefore, they may be easily shear-melted by slight mechanical disturbance e.g., by simple shaking. Manipulation of colloidal solids by external fields is very effective, offering possibilities to control the colloidal micro-structure upon re-crystallization, as well as the formation and re-juvenation of colloidal glasses [17, 38, 39, 40, 41].

Glasses appear in many classes of materials ranging from silicates over metals to polymers. It is possibly fair to state that up to now, there is no general answer to the important questions 'what defines a glass and the glass transition?', and 'what is the nature of the vitrification process?'. We here adopt a pragmatic point of view and accept any amorphous, short-range ordered solid as a glass. In particular, this definition of a glass does not rely on the presence of particular structural motifs [42]. Further it does not include stability against crystallization or the presesence of specific types of dynamics nor does it request the complete absence of long-time relaxation processes [43, 44]. Rather, we stress that in colloidal systems, crystallization and vitrification can be viewed as competing processes sometimes occurring on similar time scales [45]. Any study of colloidal glasses is therefore concerned with the practical aspect of how to avoid crystallization and instead obtain a glass. This question seems particularly interesting in the case of suspensions displaying repulsive and spherically symmetric pair interactions. The two representative colloid systems considered in this paper behave quite differently. HS systems of volume fractions of $\Phi \approx 0.57$ and larger readily vitrify from the melt [2]. Crystallization via homogeneous nucleation at such large volume fractions is observed – if at all - only on very long time scales and, typically, under a reduced influence of gravity [46, 47, 48]. By contrast, crystallization from the melt appears to be a very rapid and effective process in charge stabilized systems [49]. There, creating a colloidal glass turns out to be rather difficult. Several different strategies have been applied to obtain and stabilize the meta-stable glass state in the two systems. Topological, thermodynamic and kinetic approaches complement each other. They are conceptually distinguished with ease, but in the attempts to vitrify a given experimental system, they often appear in combination.

The present paper will review some of these investigations. We will first discuss examples of topological, thermodynamic and kinetic approaches with some focus on our own investigations. For all the given examples, the route taken into the glass state appears to be relatively clear. We then turn to a recently observed amorphous solid formed at low density from thoroughly deionized, highly charged spheres. We report some preliminary experiments demonstrating that the investigated samples display a finite shear rigidity, a short range order and a dynamical heterogeneity. Remarkably, this amorphous solid can be obtained at volume fractions as low as $\Phi \approx 0.005$. We end with some short conclusions.

**The topological approach**

We start our compilation with a special case: suspensions of charged nano-clay platelets, commonly known under the name Laponite[®TM1]. These show long-ranged dipolar and quadrupolar contributions to their pair interaction potential [50]. These systems do not form crystals. Rather, they assume an amorphous structure in the solid phase through different paths already at low volume fractions [51]. In analogy to the Wigner-crystals formed by spheres through electrostatic repulsion, such states have been termed Wigner-glass by some authors [41]. Other authors explain the arrested states as colloidal gel resulting from the partially attractive interactions. Several attempts have been made for a comprehensive interpretation of the results reported in the literature which try to reconcile the different existing views [52, 53]. The peculiar phase behavior of nano-clays can be traced back to its long ranged repulsion combined with a shorter-ranged anisotropy of the interaction potential. They therefore constitute an example of inhibited crystallization due to topological frustration.

✉     Thomas Palberg
       palberg@uni-mainz.de

---

[1] Laponite[®TM] is a trade mark of BYK Additives Ltd. In the community, commercially available nano-clay species, as well as home made particles of this synthetic mineral and models thereof, are commonly known under the generic name Laponites and purchased particles are not always correctly mentioned.

Another example is given by hard ellipsoids which can pack even more dense than hard spheres, but in experiments typically form solids with short-range order only [21]. A very famous example is the Bernal-type glass. Starting from spherical particles, tetrahedral units are formed, which cannot fill a three dimensional volume in a close packed regular fashion [54, 55, 56, 57, 58]. Mutually frustrating polyhedral units are also found in vitrifying in HS systems and have been proposed to form the structural basis for vitrification [42, 59].

HS are the most widely studied glass forming colloids. Theory, simulation and experiment fruitfully complemented each other in an impressive number of studies. There are two important differences between experimental HS approximants and their ideal computer counter parts: polydispersity and the influence of gravity. Both affect strongly the phase behavior, the phase transition dynamics and the glass transition. For strictly monodisperse HS not subject to sedimentation, the most recently determined locations of the freezing and melting volume fractions are $\Phi_F = 0.492$ and $\Phi_M = 0.545$, respectively [60]. Experimental systems are inevitably polydisperse and further may show a slightly soft repulsive pair interaction [61, 62]. In these cases some shifts of the coexistence region are expected [63, 64]. Strongly polydisperse systems will further show fractionation effects [65].

Colloid polydispersity is most conveniently expressed in terms of the polydispersity index, PI, defined as the standard deviation divided by the average size. Crystallization in moderately polydisperse experimental HS systems (PI < 0.06) proceeds via a two step mechanism [66] in which first compressed precursors of short-range order are are formed that later transform to long-range ordered crystallites. First seen in light scattering experiments this mechanism has later also been observed in simulations [67, 68, 69] and studied in detail using high resolution microscopy [70, 71]. Most importantly, this precursor mediated mechanism occurs also in very well buoyancy matched systems. Presumably, gravity enhances its effects, but does not cause it [48]. With increasing volume fraction of the melt, the measured incubation times increase drastically and the crystal nucleation rate densities drop sharply. From the perspective of particle dynamics, there is an arrest of flow and emergence of activated processes at the glass transition [45]. From the perspective of crystallization the precursor stage is significantly prolonged. Very recent investigations utilizing spatially resolved multi-speckle correlation technique on HS systems of different volume fraction could further demonstrate a close spatial and kinetic correlation of dynamical heterogeneities typical for glasses and structural heterogeneities correlated with precursor formation [72]. The more precursors nucleate and fill up the sample until they start to intersect, the less mobile the system gets on average. Finally it is simply stuck in the process of transforming to the stable crystalline state [48].

We note, that a further kinetic slowing may be observed for strongly polydisperse systems (PI > 0.07), where crystallization affords fractionation processes within or at the surface of precursors [73]. Conversely, the addition of small amounts of non-adsorbing polymer enhances the particle mobility [74, 75]. This can lead to a de-vitrification of the samples [76]. At even larger polymer content a re-vitrification as attractive glass is observed [18, 19, 77]. Added polymer, however, also can enforce crystallization. We come back to this point below.

On short time scales gravity may facilitate local jamming, while on long time scales it may induce gradients in particle density or stratification effects [78, 79]. This can be overcome by studying the samples under micro-gravity or improving the buoyancy match using micro-gel spheres [46, 47, 48]. On the other side, gravity is quite useful for a determination of the equation of state [80] or the phase behavior [78, 81, 82]. In Fig. 1, we show an image of the first colloidal glass obtained and systematically characterized [3, 7, 8, 10, 11] by Pusey and van Megen. They used suspensions of HS-like particles at elevated effective HS volume fractions $\Phi \gtrsim 0.57$. To be specific, their system comprised of polymethylmetacrylate (PMMA) spheres, sterically stabilized by 12-hydroxy-stearic acid (PHSA) and suspended in a solvent mixture of decalin and carbon disulfide to match the refractive index of particles and solvent. In Fig. 1, the three originally vitrified samples to the right showed only minimal settling under gravity. This, however, was sufficient to allow for crystal formation via heterogeneous nucleation at the top surface of the samples. The crystals are of slightly larger packing fraction and dilute the immediately adjacent suspension creating a region of enhanced mobility directly at their surface. Thus they could grow deep into the glass forming iridescent columns. Columnar growth was also observed upon expansion of an amorphous sediment formed by centrifugation [83] presumably facilitated by a similar mechanism.

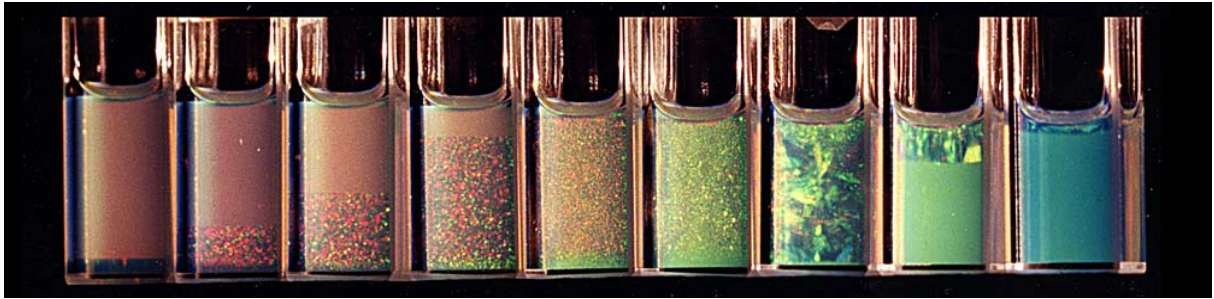

Fig. 1: (color online) The phase behavior of hard-sphere like colloidal particles. The effective HS volume fraction increases from left to right from 0.475 to 0.643. The leftmost sample shows a fluid bulk with a small amount of columnar crystal formed after gravitational settling. The next three vials show coexistence of a fluid with fcc crystals settled under gravity. Two middle samples are fully crystalline with crystallite sizes decreasing with increasing volume fraction. The three rightmost samples originally formed a colloidal glass. After heterogeneous nucleation at the slightly sedimentation-diluted top, and by subsequent growth, also these became (partially) filled with crystals. (Image courtesy W. van Megen)

Effects of gravity may combine with those of strong polydispersity enforcing fractionation [63, 64, 65] and of added polymer inducing attraction [4, 19]. But also minimized influences of gravity in attractive polydisperse systems may lead to interesting results. We recently investigated a one-component sample buoyant micro-gel spheres suspended in 1-Ethylnaphtalene [84, 85, 86]. We observed crystallization up to $\Phi = 0.593$ for the pure HS case. We then added non-adsorbing polymer (size ratio polymer/particle 0.01) to obtain polycrystalline samples below a polymer reservoir packing fraction $\eta_{res} < 0.2$ and $\Phi \geq 0.55$ but again no glass was observed. For larger $\eta_{res}$ coexistence between strongly compressed crystals ($\Phi_{Xtal} = 0.734$) and a fluid phase of $\Phi \approx 0.52$ was observed. Finally, for $\eta_{res} > 0.6$ an attractive glass or gel was found.

This complex behavior is not fully consistent with previous measurements of other groups and theoretical expectations [87, 88, 89, 90, 91, 92, 93, 94, 95, 96, 97]. Neither a repulsive glass nor a phase separation into crystal and coexisting low density fluid were observed. We attribute this behavior to a combination of several factors. Our excellent buoyancy match reduces the risk of jamming and depletion attraction further enhances particle mobility. This favors crystallization which leads to a reduction of the volume fraction of the remaining melt. Altogether, these effects suppress the repulsive glass. Moreover, fractionation leads to highly compressed crystals, but further enriches the melt by left-over odd particles. Fractionation thus renders crystallization via homogeneous nucleation incomplete and the larger part of the sample remains as only slightly diluted melt. In the absence of gravity neither crystals nor melt settle sufficiently to start the phase separation from a low density supernatant phase. The system therefore appears to be kinetically stuck on its way into equilibrium by a combination of several factors.

In the case of charged sphere systems typically smaller particles are studied, which are less prone to sedimentation. Practically all systems crystallize quickly already at low volume fractions to form body or face centered cubic crystals [9]. It is therefore interesting to check whether charged spheres can also form HS-like glasses and how in this case precursor mechanisms (recently observed in experiments at low volume fractions [70] and in simulations [98]) affect the transition to an amorphous state.

Already several decades ago, the very existence of amorphous solids from charged colloidal spheres in aqueous suspension had been shown. Often, the authors identified the glassy state from the absence of crystals upon visual inspection [99]. Others used the split of the second peak of the static structure factor [9]. We note that this split is neither a universal nor a unique feature, and it also occurs in very mobile, rapidly crystallizing melts of charged spheres and metals [100]. An example taken from [101] is given in Fig. 2. Further experimental criteria are a certain height of the principal peak of the liquid-like static structure factor and the development of a plateau in the intermediate scattering function [102, 103]. Most charged sphere systems show this transition at lower volume fractions than HS. Typical values of $\Phi_G$ range 0.2-0.4 and show a trend too increase with increasing electrolyte concentration [9, 102, 103]. This is rationalized considering that with increased charged sphere concentration an effective (self-)screening of the electrostatic repulsion is found, and the pair interaction approaches the theoretical limit of HS. While one component charged and hard silica spheres show a structural signature comparable to that of HS, they display a different evolution of the dynamics and some differences in the q-dependence of the non-ergodicity parameter [104].

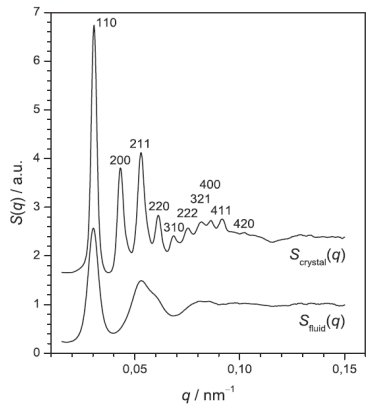

Fig. 2: Example of USAXS static structure factors of a deionized binary mixture of two strongly polydisperse silica sphere species ($a_A$ = 84,4nm; $a_B$ = 104,7nm; size ratio 0.84, mixing ratio $x_A$ = 0.4; $PI_A \approx PI_B \approx 0.09$; effective $PI_{MIX}$ = 0.12). The lower curve shows the structure factor of the melt. Note the clearly visible split in the second peak. The upper curve was shifted for clarity. It shows the structure factor of the corresponding polycrystalline bcc solid (with Miller indices indicated) which was obtained a few minutes later after complete crystallization. (Reproduced with kind permission from [101])

Furthermore, the kinetic pathway of charged spheres into the amorphous solid may proceed differently as well. Schöpe et al. studied the nucleation rate density in buoyant, refractive index matched suspensions of perfluorinated charged latex beads [105]. They observed the nucleation rate density to increase exponentially with increasing particle number density up to the largest investigated volume fractions. Observing structure and micro-structure of the obtained solids by static light scattering, they found that with increasing particle concentration the samples evolved continuously from a polycrystalline state over a nano-crystalline to an amorphous state. Assuming the size of crystallites to be always determined by cessation of growth upon intersection, they suggested that the amorphous solid had been formed by the intersection of crystallites already during nucleation, and that any further growth and coalescence was suppressed by orientational frustration. It seems that in this case rapid crystallization aids forming an amorphous solid from topologically incompatible units.

**The thermodynamic approach**

Here, we consider systems, where the colloidal fluid stays the thermodynamically stable phase and undergoes a vitrification process. Most prominent in computer simulations are Kob-Andersen-type glasses [106]. These systems were initially realized as mixtures of

Lennard-Jones particles with size ratio $\Gamma = a_S/a_L = 0.88$ and utilized to test the Mode Coupling Theory (MCT) of the glass transition [107, 108]. Size ratio and composition were chosen to mimic those found in metal-metalloid-glasses [109]. From the point of view of metal physics, such mixtures are eutectic systems, i.e. systems where the melting temperature $T_M$ is strongly decreased at the eutectic composition $x_E$. Neither compounds nor substitutional crystals are formed. Below the eutectic temperature $T_M(x_E)$, the two pure components instead segregate and form two individual crystal phases. Off $x_E$ and above $T_M$, eutectic systems crystallize only partly. Crystals of the respective majority component coexist with a melt enriched in the minority component. Occasionally in some nano-colloidal systems and recently in a strongly polydisperse binary mixture of charged spheres, the formation of Laves-phases has been observed [110, 111]. However, this phenomenon is restricted to regions close to the freezing transition and off the eutectic composition [112].

Within the eutectic gap, the liquid is the thermodynamically stable phase due to the large entropy of mixing. In HS mixtures of different size ratios, either the majority component forms its pure crystals coexisting with a minority enriched melt or compounds, pure crystal phases and melt coexist [113, 114, 115, 116]. Glasses may form at very large volume fractions above the freezing transitions for respective phases [115]. However, glasses may also form within an eutectic gap, where the fluid is thermodynamically stable [116]. In all these cases, they are again identified by their static light scattering pattern, or the development of an extended plateau in the intermediate scattering function. Fig. 3 shows an example from our investigation of a buoyant binary micro-gel HS mixture of size ratio $\Gamma = 0.785$ at eutectic composition (number ratio $x_{S,E} = 0.77$) [116]. From the fits of the intermediate scattering functions $f(q,t)$ (symbols) with theoretical expressions of mode coupling theory (MCT [117]) (solid curves), a MCT-glass transition volume fraction of $\Phi_G = 0.573 \pm 0.002$ was derived.

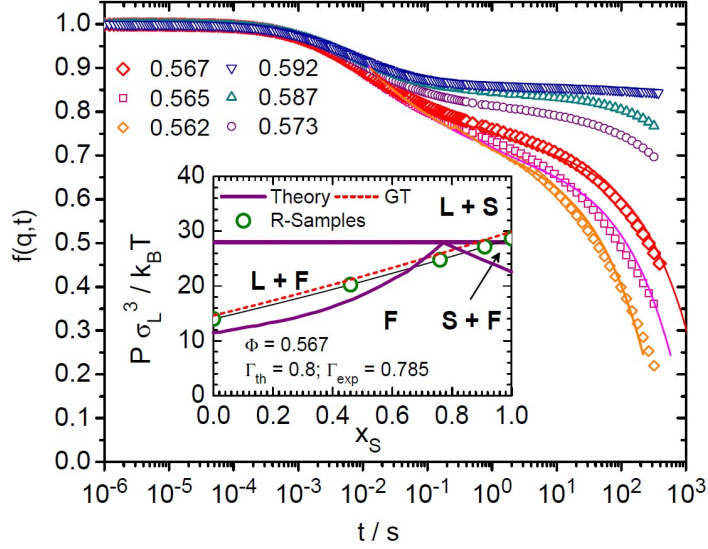

Fig. 3: (color online) Main: intermediate scattering functions f(q,t) vs. log(t) for repulsive samples of eutectic composition ($x_{S,E}$ = 0.77) with increasing $\Phi$ from bottom to top. Measurements were taken at a q-value corresponding to the main peak of the static structure factor, q = $q_{max}$ = 25.85μm$^{-1}$. Inset: binary HS phase diagram in the pressure-composition plane. $x_S$ = $n_S$/n denotes the number ratio of small spheres. F = fluid, S = crystals of small particles, L = crystals of large particles. Note that the pressure P is normalized by the cube of the large particle diameter $\sigma_L$ and the thermal energy. Thick solid lines: phase boundaries adapted from [118] for $\Gamma_{th}$ = 0.8. Thin solid line: osmotic pressure for HS mixtures of experimental size ratio $\Gamma_{exp}$ = 0.785 and volume fraction $\Phi$ = 0.567, estimated using the fundamental measure theory equation of state of [119]. Circles highlight the values for the samples of partial volume fractions $\phi_S$ = $\Phi_S/\Phi$ of 0, 0.26, 0.57, 0.83 and 1 corresponding to molar fractions $x_S$ = $\rho_S/(\rho_S + \rho_L)$ of 0, 0.47, 0.77, 0.92 and 1 (where $\rho_i$ = $\Phi_i/(4\pi/3)R_{h,i}^3$ is the particle number density of component i and $R_h$ denotes the hydrodynamic radius). Dashed line: osmotic pressures estimated the same way for a volume fraction of $\Phi_G$ = 0.573 corresponding to the mode-coupling glass transition (GT) for our repulsive systems of eutectic composition as obtained from applying a MCT analysis to the dynamic data shown in the main part. (Reproduced with kind permission from [116])

We shortly note that in binary HS mixtures the glass transition occurs for all compositions with some dependence of the transition volume fraction on the size ratio and a size ratio dependent variation with composition [120]. This explains the occurrence of glasses also in the crystalline regions of the phase diagram of binary mixtures. Although crystal nucleation is in principle possible, it is kinetically slowed as both segregation of a pure phase and compound formation require composition fluctuations. These are much slower than the density fluctuations needed for crystallization of one component HS. If vitrification is understood in terms of relaxation of a meta-stable fluid into deep local minima, then here. the glass transition kinetics are much faster than the crystallization kinetics.

For one component HS, the addition of non-adsorbing polymer shifts the melting line towards lower pressures [60], while simultaneously, the glass transition pressure is shifted upward for one component HS and binary HS mixtures [18, 19, 20, 121, 122]. In our eutectic mixture at eutectic composition and the former glass transition volume fraction, this makes the two coexisting crystal phases the thermodynamically stable state. These thermodynamic effects are aided by an increased mobility of the spheres, and together they facilitate the nucleation of fcc crystals of both components. However, crystallization is still very slow and incomplete presumably due to the required fractionation effects and a strongly increased melt-nucleus interfacial free energy (IFE) [123, 124].

Also in charged sphere systems, Kob-Andersen-type mixtures are expected to show a glass transition [125]. Several experimental studies report the observation of glasses in binary charged sphere systems. Again, glasses were identified by a split of the second peak of $S(q)$ [126], the absence of crystal Bragg reflections in reflection spectra [99], or a non-decaying intermediate scattering function [102]. However, only in the latter study of Meller and Stavans glasses were formed in a eutectic mixture.

The behavior of charged spheres differs from that of HS in particular under low salt conditions. These are reached either by exhaustive deionization in a batch procedure [99] or using advanced deionization circuits [127]. Under such conditions of a long-ranged, soft electrostatic repulsion, also azeotropic [128] or spindle type [129] phase diagrams are observed in addition to eutectic phase diagrams [82, 102]. Moreover, charged sphere phase diagrams tend show large regions of substitutional bcc alloys at low and glasses only at large volume fractions. The size ratios separating the different phase diagram types are shifted to much lower values than predicted [113, 118, 130] and observed in HS [131, 132].

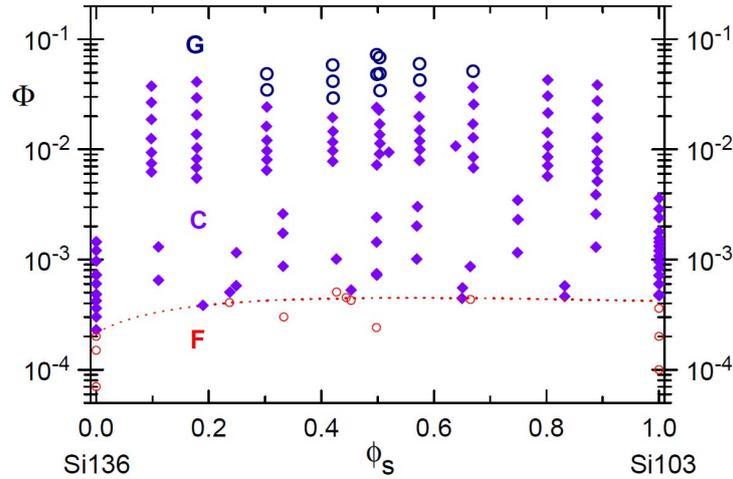

Fig. 4. (color online) Binary charged sphere phase diagram in the volume fraction – composition plane. F = fluid (small circles), C = substitutional alloy crystals of bcc structure (filled diamonds), G = amorphous solids (large circles). The phase diagram shows only a slight curvature of the freezing line (dashed). This indicates an excellent miscibility of the two components resulting in this spindle-type phase diagram. The two silica sphere species of low polydispersity (PI ≈ 0.03) and diameters of 103 nm and 136 nm, respectively, were prepared under exhaustively deionized conditions. Here the screening length is dominated by the contribution of the counter-ions released by the highly negatively charged particles. For $\phi_S$ = 0.43 corresponding to $x_S$ = 0.5 it decreases from about 2μm to about 110 nm when going from the melting density to the glass transition density. (Redrawn from [99])

In Fig. 4 we show the phase diagram of a binary mixture of charged silica spheres with size ratio Γ = 0.757. It was obtained under exhaustively deionized conditions (electrolyte concentration limited by the self dissociation product of water and the particle counter-ions). We show the Φ - $\phi_S$ plane of the phase diagram, where $\phi_S$ is partial volume fraction of the smaller spheres. To stress the influence of the potential steepness, it is instructive to compare this Fig. 4 to the inset of Fig. 3. At nearly the same size ratio as the HS eutectic mixture, the melting line of the charged system appears between $\Phi_{M,L}$ = 0.0002 for the large component and $\Phi_M$ = 0.0005 at $\phi_s$ = 0.5. This is only a very slight variation of $\Phi_M$ with composition and typical for spindle type phase diagrams. The glass transition is separated from the melting line by an extended region of bcc-crystal phase. Its lowest value is at $\Phi_G$ = 0.03 for a molar fraction of small spheres of $x_S$ = 0.5 corresponding to $\phi_s$ = 0.43. Interestingly, this coincides with that volume fraction, where the Debye screening length $\kappa^{-1}$ has decreased to values comparable to the average particle diameter. If one assumes a crude estimate of the effective HS diameter as $\sigma_{HS,eff} = \sigma_{core} + 2\kappa^{-1}$, the effective volume fraction is $\Phi_{G,eff} = 0.24$. It is thus very close to the one observed for one component system of charged latex spheres [9], silica spheres [102, 104] and of buoyancy matched perfluorinated spheres [105].

In both one-component charged sphere systems and their binary mixtures forming substitutional alloys, a crystal phase is found between the fluid and the glass. In this phase, the crystallite size decreases drastically with increasing $\Phi$ and a frustration mechanism similar to the case of one-component charged sphere systems seems to apply. The glass transition inside an eutectic gap, where it emerges from the fluid state without intervening crystal, was so far only studied at elevated electrolyte concentration under HS-like interaction conditions [102]. Further experimental studies at charged sphere size ratios suitable for the formation of eutectics [132] but at low salt concentration are therefore highly desired to complement the many HS investigations of Kob-Andersen glasses.

**The kinetic approach**

In the inset of Fig. 3 the glass transition of a binary HS mixture occurs in the stable fluid region. In Fig. 4, the glass transition of charged spheres is separated from the stable fluid by an intervening region of (rapid) crystal formation. This also applies to one component HS (c.f. Fig. 1). In several of the latter cases we have seen that the glass was reached by some variant of the topological approach. If a meta-stable solid state exists, it should, however, also be accessible via a kinetic approach. I.e. a glass is formed, if crystal nucleation is sufficiently slowed. We have mentioned that this occurs in the vitrification of HS mixtures where vitrification competes with compound formation – and wins. The same effect, presumably, is causing the vitrification of metal alloys, although this has not yet been studied in full microscopic detail [133]

For colloids, several possibilities to slow nucleation exist according to classical nucleation theory (CNT). CNT takes a macroscopic view on the microscopic phenomenon of nucleation [134, 135, 136]. Despite its still controversially discussed assumptions and range of applicability, and despite its clearly limited predictive capabilities, it nevertheless is widely believed to capture at least the essentials of crystal nucleation as a thermally activated process [49, 137, 138, 139]. The CNT expression for the nucleation rate density $J$ adapted to colloids reads [140]:

$$J = J_0 \exp\left(-\Delta G^*/k_B T\right) \qquad (1)$$

Here $J_0$ is a kinetic pre-factor depending on particle number density n and the long-time self-diffusion coefficient. The energy barrier for nucleation appearing in the Boltzmann factor is given by:

$$\Delta G^* = \frac{16\pi}{3}\frac{\gamma^3}{(n\Delta\mu)^2} \qquad (2)$$

where $\Delta\mu$ is the chemical potential difference between the fluid and the crystalline phase and $\gamma$ is the surface tension. In the literature reduced interfacial free energies (IFEs) are utilized to compare systems of different density and interaction type, $\sigma = \gamma\, n^{-2/3}$, and the thermal energy $k_B T$ is used as energy unit.

Eq. (1) and (2) reveal three different possibilities of slowing the nucleation. First, one could reduce the kinetic pre-factor. As discussed above, this route is taken in systems requiring composition fluctuations for the formation of crystals. It is also important in hairy systems consisting of block-copolymer micelles. Changing the relative block lengths, allows tuning the particle interaction between that of hard spheres and that of polymer coils. In a combined experimental and theoretical study on the dynamic phase behavior of such systems [141], it was observed that with increasing hair length as compared to the core size, the hydrodynamic radius governing diffusion increased significantly beyond the radius of the bare HS-like core. Further an increasing softness of the still spherically symmetric interaction was obtained, which influenced the phase behavior. Depending on the micellar aggregation number, the systems showed either bcc crystalline or glassy phases, both adjacent to the low volume fraction fluid phase. A similar approach seems feasible also for charged-uncharged block-copolymers in aqueous suspension, but has so far not been reported.

The second experimentally variable quantity in Eqn. (2) is the chemical potential difference $\Delta\mu$. For HS it is entropy governed and increases with volume fraction to maximum values of a few $k_B T$. For charged spheres, much larger values are obtained due to the enthalpic contribution to the free energy [142]. This restricts the region of large nucleation barriers to regions of the phase diagram very close to the fluid solid phase boundary, where $n \gtrsim n_F$. However, as discussed above, even these low $\Delta\mu$ appear to be not sufficient to suppress crystallization in the one component systems and substitutional alloy forming mixtures studied so far.

The third handle is the IFE. For HSs, equilibrium values from direct measurements and computer simulations on planar interfaces coincide with measured non-equilibrium IFEs from nucleation experiments evaluated using CNT. Typical values are in the range $\sigma_{HS}$ = (0.56–0.68) $k_BT$ [48, 60, 143, 144, 145, 146, 147, 148]. Since freezing in HS systems is driven solely by entropy, the reduced IFE is independent of volume fraction even at large meta-stabilities [149]. For charged spheres, the reduced IFE depends linearly on meta-stability. In a recently performed systematic analysis for five one-component charged sphere systems and one binary mixture, we extrapolated non-equilibrium values $\sigma(\Delta\mu)$ to obtain the equilibrium values, $\sigma_0$ [150]. Interestingly, we did not find any correlations of $\sigma_0$ to parameters determining the interaction strength and range. However, our study revealed a remarkable anti-correlation of $\sigma_0$ to the PI. $\sigma_0$ decreased from 4.28 $k_BT$ to 1.13 $k_BT$ when the PI increased from 0.025 to 0.8. This decrease was steepest at low PI. Moreover, the IFEs of the mixture lay significantly below those of the pure components. We rationalized this behavior of $\sigma_0$ by a decrease of the entropy difference between melt and adjacent crystal, as suggested by Turnbull's Rule [151, 149]. The observed behavior further suggests that the IFE of near monodisperse charged sphere samples should be still larger. In turn, and amplified by the low value of $\Delta\mu$ close to the phase boundary, the nucleation barrier of monodisperse charged spheres may become sufficiently large to efficiently decrease the nucleation rate density. We further tentatively propose that this would leave a suitable system enough time to form an amorphous solid before the onset of crystallization.

**A low density charged sphere glass**

*Experimental*

Recently, we performed a study of the phase behavior of charged spheres kindly provided by BASF, Ludwigshafen. Under deionized conditions, all except one sample crystallized readily from their meta-stable melts. This particular sample first formed a transient amorphous solid from which it finally crystallized after some time. In this section we give a preliminary account of our findings. Like all the other samples also the particles of Lab code PnBAPS118 (manufacturer Batch No. 1234/2762/6379) are a 35:65 W/W copolymer of Poly-n-Butylacrylamide (PnBA) and Polystyrene (PS). Their diameter was determined by the manufacturer utilizing dynamic light scattering to be (117.6 ±0.65) nm. This corresponds to a nominal polydispersity index of PI = 0.011. The effective charge number of the particles was determined from number density dependent conductivity measurements on deionized samples to be $Z_{eff}$ = 647±18 elementary charges. This type of effective charge corresponds to the number of freely moving counter ions, coincides well with Poisson-Boltzmann cell calculations and accounts for charge regulation and charge renormalization [36, 152].

The supplied stock suspension ($\Phi \approx 0.2$, n ≈ 230 µm$^{-3}$ = 2.3 10$^{20}$ m$^{-3}$) was first diluted and stored over mixed bed ion exchange resin (Amberlite, Rohm & Haas, France) for a few weeks under occasional gentle stirring. It was then filtered to remove dust, resin debris and coagulate, regularly occurring upon first contact with the exchange resin. The procedure was repeated twice using fresh resins and the cleaned stock solution was then stored in a fridge. From this stock samples of desired number densities (corresponding to volume fractions of 0.02 and less) were prepared by dilution in 7 ml sample vials with freshly rinsed ion exchange resin added. Samples were sealed against airborne $CO_2$ with Teflon® septum screw caps (Sigma Aldrich, Germany). They were left for more than two months until the crystallite sizes obtained after daily gentle shaking became constant, indicating thoroughly deionized conditions [99]. Experiments were performed in dependence on number density and waiting time $\tau_W$ defined as the time after last gentle shaking.

Samples were studied using a home build multi-purpose light scattering instrument [153]. It combines a static light scattering experiment to determine the static structure factor $S(q,\tau_W)$ with a dynamic light scattering experiment to measure the normalized intensity autocorrelation function $g^{(2)}(q, t, \tau_W)$. Its main features are a counter-propagating collinear illumination by a split laser beam (Innova 70C-Spectrum, Coherent, Santa Barbara, CA) Spectra Physics) and a two-arm goniometer. The opposing illumination stages and the opposing detection arms carry optimized illumination and detection optics, respectively. This allows performing static and dynamic experiments quasi-simultaneously under the same scattering vector and on the very same sample, without the need to move the fragile solids between different experimental set-ups.

The static side illumination is further used for torsional resonance spectroscopy (TRS) to determine the shear modulus G of the sample. In TRS the sample cell is set into low-amplitude oscillations about its vertical axis, which excites the eigenfrequencies of the solid in the known geometry. A reference signal is obtained from the reflection of a laser beam off a small mirror fixed to the sample outside. A second laser beam illuminates the sample and is scattered upon a position sensitive detector (PSD, SSO-DL100-7, Silicon Sensor, Berlin, Germany). For crystalline samples, an individual Bragg reflection is chosen and its peak position change is detected as a function of time. For amorphous samples, a scattering vector $q$ on the low-$q$ slope of the primary peak in $S(q)$ is selected. The periodic change in the scattered light intensity $I(q, t)$ is recorded using the PSD in integral mode. Using a dual channel lock-in amplifier (SR530, SRS, Sunnyvale, CA) the resonance spectrum is recorded for frequency intervals of (0.5-10) Hz. The position of the eigenfrequencies then allows determining the shear modulus $G(n, \tau_W)$ of the sample [36].

*Results*

Images of some representative samples are shown in Fig. 5. Vial height is approximately 4cm. The particle concentration increases from left to right. With increasing concentration, one recognizes an amorphous structure without any crystallite Bragg reflections, a partially crystalline structure with columnar crystals, two polycrystalline samples and a nano-crystalline sample, respectively. The short-range order in the vial at lowest concentration is very pronounced. This can be seen from the two rainbow-like scattering patterns with pronounced blue regions. In this white light scattering experiment under fixed observation angle, changes in scattering vector translate into different scattered wavelengths. Thus, the clear color separation results from the pronounced and narrow first two peaks in the fluid-like $S(q)$. This image was taken a few hours after last shaking. In this image, the sample is identified to be solid by the non-sedimented ion exchange resin splinters marked by the arrows. After 5 days columnar crystals have grown into the glassy bulk after nucleation at the cell wall.

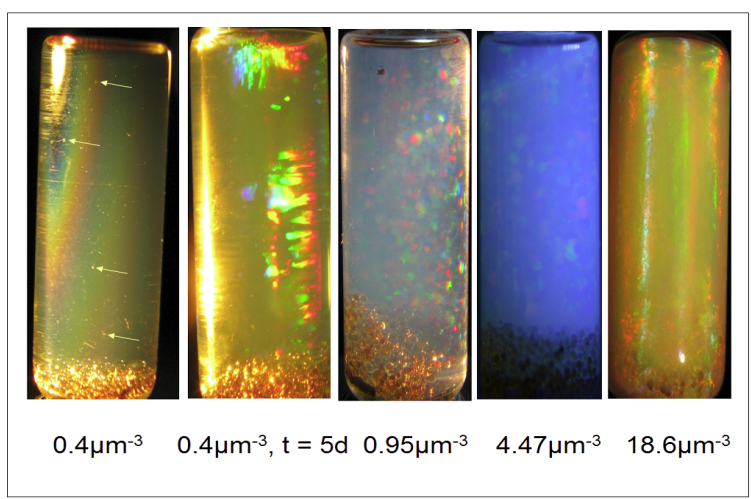

Fig. 5: (color online) Photographs of PnBAPS118 samples taken at different number densities, n, as indicated. Sample height is ca. 4 cm. The leftmost sample is an amorphous solid, as recognized from the pronounced double rainbow and the non-sedimented ion

exchange resin splinters marked by the arrows. After five days columnar crystals have grown into the amorphous solid after heterogeneous nucleation at the cell wall. At larger number densities homogeneously nucleated crystals are obtained. Crystallite sizes decrease with increasing n, yielding polycrystalline and nano-crystalline solids. Moreover, crystallization occurs on ever shorter time scales down to minutes.

The appearance of the samples changes with time in a number density dependent way. In general, the time scale of vitrification decreases, as does the time scale of crystallization. Effects of sedimentation become discernible after several weeks. The interval between solidification and crystallization shrinks with increasing n. Therefore, transient amorphous solids could be identified by visual inspection only for $0.35$ $\mu m^{-3} \leq n \leq 12$ $\mu m^{-3}$ ($3.4 \cdot 10^{-4} < \Phi < 0.01$). Above, samples crystallize too quickly to be unambiguously identified as an amorphous solid. Below, vitrification is too slow to hinder complete sedimentation of visible objects. A convenient time window for accurate instrumental measurements on the fully developed amorphous solid exists for samples of number densities $0.4$ $\mu m^{-3} < n < 8$ $\mu m^{-3}$ ($3.4 \cdot 10^{-4} < \Phi < 7.8 \cdot 10^{-3}$). After transfer from the shelf to the light scattering apparatus, these solidify into the glassy state on acceptable time scales, but crystallize slow enough to complete the measurements. Therefore $n_{GT} \leq 0.4$ $\mu m^{-3}$ is the present upper bound for the lowest number density from which PnBAPS118 forms amorphous solids. A lower bound may be estimated as $0.15$ $\mu m^{-3} < n_{GT}$ ($1.3 \cdot 10^{-4} < \Phi_{GT}$) from the largest number density, for which after one day of waiting no two step decay of $g^{(2)}(q,t,\tau_W)$ was observed.

Samples of $0.4$ $\mu m^{-3} < n < 8$ $\mu m^{-3}$ melt upon transference to the light scattering set up, but vitrify and crystallize within well accessible times. In Fig. 6, we show the static structure factor $S(q, \tau_W)$ measured at $n = 0.8$ $\mu m^{-3}$ for two different times after shaking. The height of the first peak at $\tau_W = 30$ min is about 2.8. It increases further to values around 3.5 and the peak width appears to narrow over the first few hours. First crystallites become visible after one day and after two days the system has crystallized into a bcc phase. Furthermore, the pronounced low-q intensity has disappeared for the crystal $S(q)$ and thus is neither an instrumental artifact nor caused by dust. Its appearance in the early $S(q)$ during the formation of the amorphous state could therefore indicate the presence of some large scale density fluctuations. Therefore, we will in future investigate these samples also by small angle light scattering utilizing the small angle scattering experiment of Beyer [84] in a version modified for investigations on water based samples.

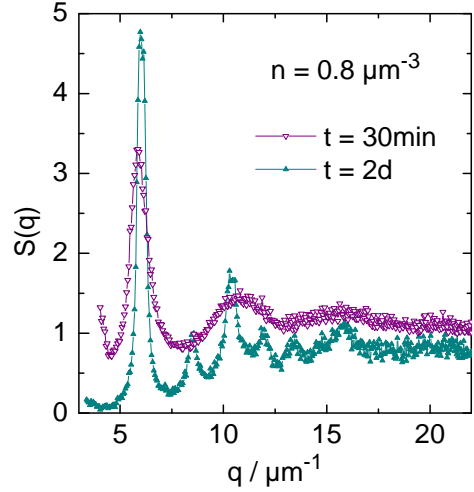

Fig. 6: (color online) Static structure factor $S(q,\tau_W)$ of a sample of PnBAPS118 at a number density of n = 0.8 µm$^{-3}$ ($\Phi$ = 6.9 10$^{-4}$) for two different waiting times as indicated.

Our means of determining the intermediate scattering function for our samples are limited, because they are both non-ergodic and show a non-negligible amount of multiple-scattering. This is clearly seen for the sample with n = 4.47 µm$^{-3}$ ($\Phi$ = 3.9 10$^{-3}$) in Fig. 4, but even present for the sample at n = 0.4 µm$^{-3}$ ($\Phi$ = 3.4 10$^{-4}$). We therefore only recorded the intensity autocorrelation function for which multiple scattering causes a very fast short time decay [154]. We attempted to integrate out these fast fluctuations by choosing the shortest sample time to be 0.1 µs and the shortest lag time of the correlator to be 1 µs. This removes the initial decay, but renders the intercept and plateau height ill defined. Still the obtained $g^{(2)}(q, t)$ can be qualitatively inspected for the presence of a second non- or slowly decaying component. This is shown in Fig. 7 for the sample at n = 0.4 µm$^{-3}$ and two different $\tau_W$ with the scattering vector coinciding with the position of the main peak of S(q). A fast decorrelation process is seen on a time scale of a few hundred micro-seconds, a second one on the time scale of some tens of seconds. Within 20 min, the second decay shifts to a few minutes. This behavior becomes more pronounced at larger waiting times. The behavior at larger n is qualitatively similar, but the evolution of the plateau occurs somewhat faster. A more systematic investigation of the evolution of the intensity autocorrelation function with $\tau_W$ is under way and will be published elsewhere. Future experiments will utilize a two-color cross correlation instrument to selectively study singly-scattered light [155].

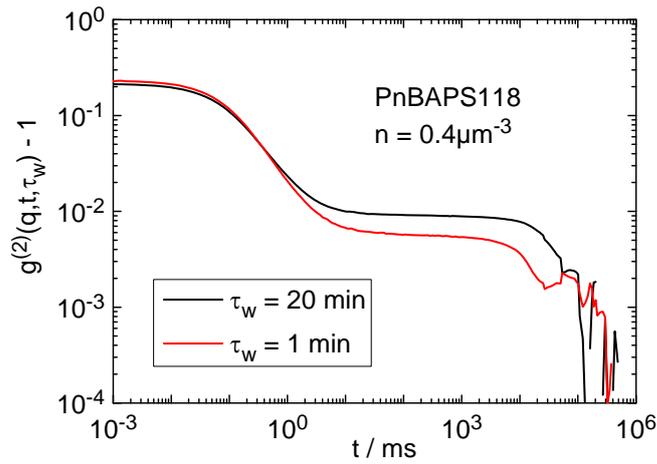

Fig. 7: (color online) Double-logarithmic plot of the normalized intensity autocorrelation function obtained for a sample at n = 0.4μm$^{-3}$ (Φ = 3.4 10$^{-4}$) and two different waiting times as indicated. Note the low intercept stemming from arbitrarily setting the shortest lag time to 0.1μs, integrating out the fast intensity fluctuations due to multiple scattering. Two relaxation processes with time constants of a few hundred micro-seconds and of a few minutes can be discriminated. With increasing waiting time, the second relaxation appears to increase.

To demonstrate solidification, we utilize TRS. A typical resonance spectrum recorded after solidification into the amorphous state at n = 4.47μm$^{-3}$ (Φ = 3.9 10$^{-3}$) is shown in Fig. 8. The resonance frequencies are identified by position of response amplitude maxima and a phase lag of 90°. This particular sample has a shear modulus of G = 0.41 Pa. Values of the shear modulus of the transient amorphous solids are slightly larger than those of the later emerging polycrystalline solids. The shear modulus is roughly proportional to the number density. This leads to a peculiar elastic behavior shortly after shaking for samples of number densities of 0.4 μm$^{-3}$ < n < 0.8 μm$^{-3}$ (3.4 10$^{-4}$ < Φ < 7.8 10$^{-3}$). The spectra can be recorded some way up the slope of the first maximum of the resonance spectrum. Then, locking is lost abruptly. This is attributed to a self-destruction of the fragile solid as it gets into resonant vibration. After a sufficiently long time of standing mechanically undisturbed, resonance spectra could be recorded also for the low density amorphous solids. The obtained shear moduli were 0.034 Pa at n = 0.4μm$^{-3}$ and 0.02 Pa at n = 0.2μm$^{-3}$. These values are close to the sensitivity limit of the present instrument and currently still contain a systematic uncertainty of about 20%. Within this error no change of G with time could be found. Systematic experiments utilizing an improved geometry of the cell will be reported elsewhere.

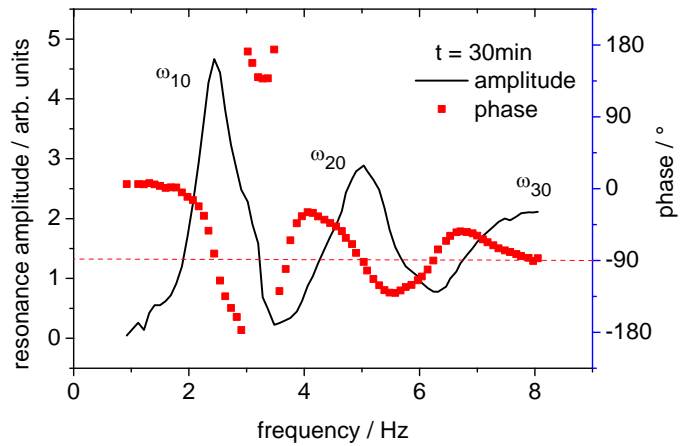

Fig. 8: (color online) Resonance spectrum of an amorphous sample of PnBAPS118 with n = 4.47 μm$^{-3}$ (Φ = 3.9 10$^{-3}$) recorded using TRS. Three resonances are clearly identified with the mode indices indicated [37]. The corresponding shear modulus is G = 0.41 Pa.

*Discussion*

The above shown data are preliminary. We believe, however, that we have clearly demonstrated the existence of meta-stable amorphous solids forming in exhaustively deionized, low number density, aqueous colloidal suspensions of highly charged spherical particles. These are characterized by a fluid-like short range order, exhibit a small but finite shear rigidity and display a plateau and a second slow relaxation process in the intensity autocorrelation function.

The nature of this solid and the details of its formation are not understood and provide a challenge for future work. One-component 3D Wigner glasses formed from spherically symmetric particles have been theoretically predicted and studied at large volume fractions, where they are stabilized by packing constraints [156]. They were further studied at low volume fractions where they form due to the long ranged electrostatic repulsion [157, 158, 159, 160 ]. The glasses formed here were observed at volume fractions as low as Φ = 0.004 under exhaustively deionized conditions. Possibly, therefore, PnBAPS118 is a candidate for a low density Wigner glass made from spherical particles.

The above result was surely unexpected. What is particularly puzzling is the fact that none of the many similar polymer or silica charged sphere colloids investigated before has ever shown a similar behavior. In view of our discussion presented above, two routes into the amorphous state seem conceivable for PnBAPS118. Both are connected to the location of the glass transition at very low volume fractions close to the number densities for freezing and melting. There, the formation of an amorphous solid could be connected to the formation of precursor structures [70, 98]. Assuming an IFE on the order of a few $k_BT$ at very low meta-stability, and combining it with the small $\Delta\mu$ close to the fluid crystal phase transition, may lead to a large nucleation barrier and hence a large size of the critical radius. One may therefore speculate, that over time, sufficiently large precursors form and jam before getting critical. An experimental indication of this may be visible in the low-q behavior of S(q) in the glassy state. This mechanism would constitute a low density analogue of the well documented topological frustration at large densities and volume fractions. As it does not require any system specific assumptions, it should in principle also be present in other systems.

The other tentative explanation assumes that the amorphous solid can be formed directly from the fluid [159] and that the barrier for the competing nucleation of crystals is sufficiently large to suppress even the formation of precursor structures. This kinetic approach could be aided by a low PI increasing the IFE. PnBAPS118 has a low nominal PI given by the manufacturer of PI = 0.011. We could, however, not yet confirm this value in our own careful dynamic light scattering experiments following [161]. The observed PI values are also low, but with a large systematic uncertainty typical for this technique in the limit of small PIs. Alternative measurements utilizing negative staining electron microscopy and small angle x-ray scattering are under way. It is therefore too early to claim a particularly low PI to be the reason for the peculiar behavior of PnBAPS118. However, the pronounced anti-correlation of $\sigma_0$ to the PI clearly warrants further investigations of systems with very low polydispersity.

Finally, one may also suspect a chemistry based possibility like a patchiness of the surface charge caused by a micro-phase separation of the two polymers. In this case, an anisotropy in the interaction potential might result and a nano-clays-like situation would be created [50, 51, 52, 53]. However, in this case the observed crystallization from the transient glass seems very unlikely to occur.

**General Conclusions**

In this paper we have reviewed several experimental approaches to complement theoretical investigations of hard and charged sphere glasses. Our overview shows that a general classification of the approaches is feasible. One may crudely discriminate topological, thermodynamic and kinetic approaches. In several cases the route taken into the amorphous state could clearly be identified. We have, however, also seen that in many experimental situations a clear distinction is difficult, in particular, when solidification is additionally influenced by gravity and polydispersity. Topological approaches rely on the incompatibility of structural units. The incompatibility may occur on different length scales ranging from individual particles to pre-critical crystal nuclei. The thermodynamic approach is most clearly demonstrated in the eutectic gap of a Kob-Andersen-type binary mixture. There the glass forms from the thermodynamically stable fluid. Glasses formed in the crystalline regions of binary mixtures appear to form either via the topological route (at large $\Phi$ or n) or the kinetic route (exploiting the slowness of composition fluctuations necessary for crystal formation). For charged spheres, so far no glasses formed from a thermodynamically stable melt have been reported. The kinetic approach is realized e.g. via a decreased pre-factor of nucleation kinetics. This is seen for compound forming and crystal phase separating eutectic mixtures. It further seems to be important in the case of sterically stabilized micelles, while the case of low density charged sphere glasses is not yet settled.

We have further demonstrated the existence of colloidal glasses close to the freezing transition of thoroughly deionized, highly charged spheres. Our future work will focus on a further characterization of PnBAPS118 glasses through systematic measurements of its structural, dynamic and elastic behavior in dependence on number density, background electrolyte concentration and waiting time. We also will revisit other charged sphere systems at low density prepared in the way PnBAPS118 has been treated. We anticipate that both will help in clarifying the particular mechanism behind this unexpected finding. Moreover, we hope, that the presented data will stimulate theoretical interest in the possibility and characteristics of one component Wigner glasses in low density charged sphere systems.

**Acknowledgements**

We are pleased to thank W. van Megen, M. Franke, S. Golde, H. J. Schöpe, J. Horbach, P. Wette and D. Herlach for fruitful discussions and permissions to reprint their data. Special thanks go to W. van Megen for providing an original image of the PMMA system. We thank BASF, Ludwigshafen for the kind gift of particles. Financial support of the DFG (Grant Nos. Ba1619/2, He1602/24, Pa459/13, Pa459/16, and Pa459/17) is gratefully acknowledged.